\documentclass[prl,aps,twocolumn,preprintnumbers, showpacs, nofootinbib,superscriptaddress,notitlepage]{revtex4-1}
\usepackage{amssymb, amsthm}
\usepackage{amsmath}    % subequations
\usepackage{mathrsfs}   % \mathscr
\usepackage{graphicx}   % figures
\usepackage{color}      % color is used in text
\usepackage{footnote}   % footnote in tabular environment, must be loaded after color
\usepackage{mathtools}
\usepackage{subfiles}

\usepackage{slashed}    % Feynman slash
\usepackage{subfigure}  % side-by-side figures
\usepackage{hyperref}   % hypertext links, including those to external documents and URLs

\usepackage{rotating}   % to rotate tables
\usepackage{multirow}   % multicolumn and multirow
\usepackage[normalem]{ulem} % strickthrough
\usepackage[utf8]{inputenc}
\usepackage{overpic}
\usepackage{subfigure}

\newcommand{\beq}{\begin{equation}}
\newcommand{\eeq}{\end{equation}}

\newcommand{\nn}{\nonumber}

\def \be  {\begin{equation}}
\def \ee  {\end{equation}}
\def \ba  {\begin{eqnarray}}
\def \ea  {\end{eqnarray}}
\def \baa {\begin{eqnarray*}}
\def \eaa {\end{eqnarray*}}
\def \lab #1 {\label{#1}}

\def\d{\hbox{{d}\kern-.20em\hbox{l}}}

\def \matrix #1 {\left(\begin{array}{cc} #1 \end{array}\right)}

\def\II{\hbox{{1}\kern-.25em\hbox{l}}}
\def \Li{\mathop{\rm Li}\nolimits}

\allowdisplaybreaks

\begin{document}

\hfill \small\textmd{TUM-HEP-1548/25}

\vspace*{6mm}

\title{Extracting Meson Distribution Amplitudes from Nonlocal Euclidean Correlations\\ at Next-to-Next-to-Leading Order}

\author{Yao Ji}
\affiliation{School of Science and Engineering, The Chinese University of Hong Kong, Shenzhen 518172, China}
\affiliation{Physik Department, TUM School of Natural Sciences, Technische Universit\"at M\"unchen, James Franck-Stra\ss e 1, D - 85748 Garching, Germany}

\author{Fei Yao}
\email{Corresponding author: fyao@bnl.gov}
\affiliation{Physics Department, Brookhaven National Laboratory,
Upton, New York 11973, USA}
%\affiliation{Center of Advanced Quantum Studies, Department of Physics, Beijing Normal University, Beijing 100875, China}
\affiliation{School of Science and Engineering, The Chinese University of Hong Kong, Shenzhen 518172, China}

\author{Jian-Hui Zhang}
\email{Corresponding author: zhangjianhui@cuhk.edu.cn}
\affiliation{School of Science and Engineering, The Chinese University of Hong Kong, Shenzhen 518172, China}
%\affiliation{Center of Advanced Quantum Studies, Department of Physics, Beijing Normal University, Beijing 100875, China}

\begin{abstract}
We present the first complete result for the next-to-next-to-leading order (NNLO) hard matching kernel indispensable for a precision extraction of 
light meson distribution amplitudes from lattice calculations of equal-time nonlocal Euclidean correlation functions. The results are given in both coordinate and momentum space, with the renormalization and matching accomplished in a state-of-the-art scheme. Our results can be used in both large-momentum effective theory and short-distance factorization approaches. Notably, our coordinate space kernel
is directly applicable to nonsinglet quark unpolarized and helicity generalized parton distributions as well. We also illustrate the numerical impact of the NNLO matching, using the pion distribution amplitude as an example.
\end{abstract}

\maketitle

%%%%%%%%%%%%%%%%%%%%%%%%%%%%%%%%%%%%%%%%%%%%%%%%%%%%%%%%%%%%%%%%%%%%%%%%%%%%%%%%%%%
%\section{Introduction}
%\label{SEC:Introduction}
%%%%%%%%%%%%%%%%%%%%%%%%%%%%%%%%%%%%%%%%%%%%%%%%%%%%%%%%%%%%%%%%%%%%%%
%%%%%%%%%%%%%%%%%%%%%%%%%%%%%%%%%%%%%%%%%%%%%%%%%%%%%%%%%%%%%%%%%%%%%%
{\it Introduction: } The lightcone distribution amplitudes (LCDAs) of mesons encode important nonperturbative information about their internal structure and play an essential role in the description of hard exclusive processes involving large momentum transfer~\cite{Beneke:1999br,Beneke:2001ev}. Such processes are important for determining the fundamental parameters of the Standard Model and searching for new physics. Typical examples include $B\to \pi l \nu, \eta l \nu$ giving the CKM matrix element $|V_{ub}|$, $B\to D \pi$ used for tagging, and $B\to \pi\pi, K\pi, K\bar K, \pi\eta, \dots$ which are important channels for measuring CP violation (see, for instance~\cite{Stewart:2003gt,Buchalla:2008jp,Belle-II:2018jsg,Cui:2022zwm,Lu:2022kos} and references therein for details). 

Among all LCDAs, the leading-twist LCDAs of light mesons --- such as the pion and the kaon --- are the simplest and most extensively studied. They provide a probability amplitude interpretation of how the longitudinal momentum of the parent meson is distributed among its quark constituents in the leading Fock state~\cite{Lepage:1979zb}. However, as LCDAs are intrinsically nonperturbative quantities, reliably calculating them from QCD theory poses a significant challenge. Over the past few decades, various phenomenological analyses have been carried out to determine light meson LCDAs~(see, e.g., \cite{RuizArriola:2002bp,Chang:2013pq,Stefanis:2020rnd}). Recently, substantial progress has also been made towards computing the radiative corrections in photon-pion reactions at higher orders~\cite{Gao:2021iqq, Braun:2021grd, Chen:2023byr, Ji:2024iak}, which is necessary to achieve a precise determination of the leading-twist pion LCDA from upcoming experimental data at the Electron-Ion Collider and Belle II. Meanwhile, lattice QCD efforts on accessing their lowest few moments~\cite{Gockeler:2005jz,Braun:2006dg,Boyle:2006pw,Arthur:2010xf,Braun:2015axa,Bali:2017ude,RQCD:2019osh,Pang:2024kza} through operator product expansion (OPE) have provided valuable complementary information.

In the past few years, significant advancements~(see \cite{Cichy:2018mum,Ji:2020ect,Constantinou:2020hdm,Constantinou:2022yye} for a recent review) have also been made in extracting the full LCDAs of light mesons, rather than just their moments, from lattice simulations of nonlocal Euclidean correlation functions~\cite{Liu:1993cv,Braun:2007wv,Ji:2013dva,Ji:2014gla,Ma:2017pxb,Lin:2017snn,Radyushkin:2017cyf,Detmold:2005gg,Chambers:2017dov}. In particular, the equal-time quark correlations defined on a Euclidean space interval, also known as the quasi-light-front (quasi-LF) correlations, have been widely employed. Their Fourier transform defines the so-called quasi-DAs, which, after a proper nonperturbative renormalization, can be connected to LCDAs through a large-momentum effective theory (LaMET) factorization~\cite{Ji:2013dva,Xiong:2013bka,Ji:2020ect}. At short distances, the quasi-LF correlations can also be linked to LF correlations defining the LCDAs through a short distance factorization in coordinate space~\cite{Radyushkin:2017cyf}.

Based on the methodology above, several lattice calculations of light meson LCDAs have been conducted~\cite{Zhang:2017bzy,Bali:2017gfr,Zhang:2017zfe,Bali:2018spj,Zhang:2020gaj,LatticeParton:2022zqc,Baker:2024zcd,Cloet:2024vbv}, with the accuracy being limited to the next-to-leading order (NLO), as the perturbative hard matching kernel in the factorization formula is currently available only up to the NLO~\cite{Ji:2015qla,Liu:2018tox,LatticeParton:2022zqc,Yao:2022vtp,Ma:2022ggj}. While higher-order corrections are essential for reducing systematic uncertainties, they remain unexplored for LCDAs. In this work, we present, for the first time, the next-to-next-to-leading order (NNLO) matching kernel required for the lattice extraction of light meson LCDAs from quasi-LF correlations (for relevant advances in unpolarized isovector quark parton distribution function (PDF), see Refs.~\cite{Li:2020xml,Chen:2020ody,Cheng:2024wyu}). Our results are given in both coordinate and momentum space, making them applicable to both short distance factorization and LaMET approaches. We also investigate the numerical impact of the NNLO matching in both coordinate and momentum space, using the pion LCDA as an example.

{\it Theoretical framework:} Our starting point is the following quark quasi-LF correlation 
\begin{align}\label{eq:quasilfcorrl}
\tilde h(z_{12},\lambda=z_{12}\cdot P)&=\langle 0|O_q(z_1,z_2)|M(P)\rangle,\nn\\
O_q(z_1,z_2)&=\bar\psi(z_1)\Gamma [z_1,z_2]\psi(z_2),
\end{align}
where $|M(P)\rangle$ denotes the external light meson state with momentum $P^\mu=(P^0,0,0,P^z)$, $\lambda$ is the quasi-LF distance. $\Gamma=\gamma^z\gamma_5$, $z_1, z_2$ are 4-vectors along the spatial $z$ direction, and $[z_1, z_2]$ represents a straight Wilson line in the fundamental representation that ensures gauge invariance of the operator, 
\beq
[z_1, z_2]= {\cal P} \exp\left[ig\int_{0}^{1}dt\, z_{12}\cdot A(z_2+ t z_{12})\right]
\eeq
with $z_{12}^\mu=z_1^\mu-z_2^\mu={{z}}_1 v^\mu-{{z}}_2 v^\mu={{z}}_{12} v^\mu$ ($v^2=-1$) and ${\cal P}$ indicating the path-ordering. %In Eq.~(\ref{eq:quasilfcorrl}) we have multiplied the r.h.s. with a translation phase factor such that $\tilde h$ depends on the separation $z_{12}$ only.

The quark bilinear operator $O_q(z_1,z_2)$ in Eq.~(\ref{eq:quasilfcorrl}) has been shown to renormalize multiplicatively~\cite{Ji:2017oey,Green:2017xeu,Ishikawa:2017faj}. As a result, the ultraviolet (UV) divergences in $\tilde h(z_{12},\lambda)$ can be removed by dividing by the rest frame matrix element of $O_q(z_1,z_2)$ (denoted as $\tilde h(z_{12},\lambda=0)$) as
\beq\label{eq:ratiorenorm}
\tilde h_R^{\rm ratio}(z_{12},\lambda)=\frac{\tilde h(z_{12},\lambda)}{\tilde h(z_{12},\lambda=0)},
\eeq
where the subscript $R$ denotes renormalized quantities, and the superscript indicates the renormalization scheme. This is known as the ratio scheme~\cite{Orginos:2017kos} in the literature, and is a viable renormalization scheme at short distances %In the equation above, we have used the subscript $R$ to denote renormalized quantities.
where $|z_{12}^2|\ll 1/\Lambda_{\rm QCD}^2$. Note that $\tilde h_R^{\rm ratio}(z_{12},\lambda)$ is independent of the regularization scheme. The renormalized quasi-LF correlation can then be related to the LF correlation defining the LCDAs via the following short distance factorization formula~\cite{LatticeParton:2022zqc,Yao:2022vtp}
\begin{align}\label{eq:coordfact}
\tilde h_R^{\rm ratio}(z_{12}, \lambda) & =
\int_0^1 \!\!\! d\alpha d\beta \;C(\alpha,\beta,z_{12}^2\mu^2)h_R^{\overline{\rm MS}}(\alpha,\beta,\lambda,\mu)+ h.t., %\mathcal{O} \left(\Lambda^2_{QCD}z^2,M^2 z^2\right),
\end{align}
%
%\begin{align}
%\tilde h_R(z_{12},\lambda)=\int_0^1 du\, C(z_{12}^2\mu^2,u)h(\mu, %u\lambda)+h.t.,
%\end{align}
where $\alpha, \beta$ are Feynman parameters, $h.t.$ stands for higher-twist contributions of {$\mathcal{O}(z_{12}^2\Lambda_{\rm QCD}^2)$}, $C=C^{(0)}+a_s C^{(1)}+a_s^2 C^{(2)}+\cdots$ with $a_s=\alpha_s/(4\pi)$ represents the perturbative hard matching kernel which is known up to the NLO both in coordinate~\cite{LatticeParton:2022zqc,Yao:2022vtp} and momentum space~\cite{Ji:2015qla,Liu:2018tox,Ma:2022ggj,Yao:2022vtp}. 

It is worth pointing out that Eq.~(\ref{eq:coordfact}) follows from 
the factorization at the operator level~\cite{Yao:2022vtp}
%\begin{align}\label{eq:opefact}
%	O_q(z_1,z_2)&= \int_0^1 d\alpha d\beta\, \Big[{\mathcal{C}}(\alpha,\beta,z_{12}^2\mu^2)O_q^{l.t.}(z_{12}^\alpha, z_{21}^\beta)\notag\\
%    &+\widetilde{{\mathcal{C}}}(\alpha,\beta,z_{12}^2\mu^2)O_q^{l.t.}(z_{21}^\alpha, z_{12}^\beta)\Big],
%\end{align}}
\begin{align}\label{eq:opefact}
	O_q(z_1,z_2)&= \int_0^1 d\alpha d\beta\,C^{\overline{\rm MS}}(\alpha,\beta,z_{12}^2\mu^2)O_q^{l.t.}(z_{12}^\alpha, z_{21}^\beta),
\end{align}
which is naturally implied by the existence of local OPE. In the above equation, $z^\alpha_{12}=\bar{\alpha}z_1+\alpha z_2$ with $\bar{\alpha}=1-\alpha$. $l.t.$ stands for the leading-twist projection of the nonlocal %correlator
operator, which acts as the generating function of leading-twist local operators~\cite{Balitsky:1987bk,Anikin:1978tj, Muller:1994ses, Balitsky:1990ck, Geyer:1999uq, Braun:2018brg}. It enjoys the same definition as the quasi-LF operator in Eq.~\eqref{eq:quasilfcorrl} but with $z_{12}^2=0$.  To simplify our calculation, we set, without loss of generality, $z_1=0$ and $z_2=z$ in the discussion below. Then, the LF correlation defining the LCDA, which enters the RHS of Eq.~\eqref{eq:coordfact}, is related to the matrix element of $O_q^{l.t.}(\alpha z, \bar\beta z)$ as follows
\beq
h(\alpha,\beta,\lambda=-z\cdot P,\mu)=\langle 0|O_q^{l.t.}(\alpha z, \bar\beta z)|M(P)\rangle.
\eeq
Since Eq.~(\ref{eq:opefact}) holds at the operator level, it also applies to collinear PDFs and GPDs defined by the same operator, provided that appropriate kinematics corresponding to these quantities is taken. In other words, %in coordinate space the calculation is universal.
our coefficient function in coordinate space is universal.

In LaMET factorization, one has to resort to a different renormalization scheme. This is because, in LaMET one deals with the Fourier transform of the quasi-LF correlation, known as the quasi-DA,
%To be more specific, the quasi-DA is defined as
\beq
\tilde\phi_R(x,P^z)=\frac{1}{if_M P^z}\int\frac{d\lambda}{2\pi}e^{-i x \lambda}\tilde h_R(z,\lambda),
\eeq
where $f_M$ is the decay constant of the light meson. The quasi-DA requires correlation information at large distances. In the above equation, we haven't specified any particular renormalization scheme yet. In order that the renormalization factor does not introduce additional nonperturbative effects at large distances, we adopt the hybrid scheme~\cite{Ji:2020brr}. In this scheme,  
the ratio renormalization is used at short distances, while the Wilson line self-renormalization~\cite{Chen:2016fxx,LatticePartonLPC:2021gpi} is employed at large distances. This leads to the following result in the continuum
\begin{align}
\tilde h_R^{\rm hyb}(z,\lambda)&=\tilde h_R^{\rm ratio}(z,\lambda)\theta(z_s-|z|)
+\tilde h_R^{\overline{\rm MS}}(z,\lambda)\theta(|z|-z_{s}),
\end{align}
where $z_{s}$ is a truncation point introduced to separate short and long distances and has to be chosen within the perturbative region. 
The hybrid scheme has been implemented in the state-of-the-art lattice calculation of light meson LCDAs~\cite{Zhang:2020gaj,LatticeParton:2022zqc,Baker:2024zcd,Cloet:2024vbv} as well as other partonic quantities~\cite{Hua:2020gnw,Gao:2021dbh,Gao:2022iex,Gao:2022uhg,Gao:2023ktu,LatticeParton:2022xsd,Holligan:2023jqh,LatticeParton:2024vck,Holligan:2024wpv,Chen:2024rgi}.

The renormalized quasi-DA can be factorized into the LCDA through the following formula~\cite{Ji:2015qla}
\beq\label{eq:factorization_mom}
\tilde\phi_R(x,P^z)=\int_0^1 dy\, {\mathbb C}\big(x,y,{{\frac{\mu}{P^z}}}\big)\phi_R^{\overline{\rm MS}}(y,\mu)+h.t.,
\eeq
where $h.t.$ stands for power-suppressed contributions of $O\big(\frac{\Lambda_{\rm QCD}^2}{x^2 (P^z)^2},\frac{\Lambda_{\rm QCD}^2}{(1-x)^2 (P^z)^2}\big)$, and ${\mathbb C}$ is related to the hard matching kernel in Eq.~(\ref{eq:coordfact}) by the following Fourier transform~\cite{Braun:2021aon,Braun:2021gvv},
{
\begin{align}\label{eq:FT_kernel}
{\mathbb C}\big(x,y,\frac{\mu}{P^z}\big)=&\int \frac{d\lambda}{2\pi}  e^{-ix\lambda}  \int_0^1 d\alpha d\beta\notag\\
& \quad \times e^{i( \alpha \bar{y} + \bar{\beta} y )\lambda} C\left(\alpha,\beta, \frac{\mu^2 \lambda^2}{P_z^2}\right).
\end{align}
}

{Alternatively, the factorization formula can be organized following the spirit of Refs.~\cite{Ji:2024oka,Ji:2020ect} as
\beq\label{eq:factorization_mom_EFT}
\phi_R^{\rm \overline{MS}}(x,\mu)=\int dy\, {\tilde{\mathbb C}}\big(x,y,{{\frac{\mu}{P^z}}}\big){\tilde\phi}_R(y,P^z)+h.t.,
\eeq
where $\tilde{\mathbb C}$ is related to $\mathbb C$ as follows:
\beq
\int_{-\infty}^\infty dy\, \tilde{\mathbb C}\big(x,y,{{\frac{\mu}{P^z}}}\big)\mathbb C\big(y,z,{{\frac{\mu}{P^z}}}\big)=\delta(x-z),
\eeq
and up to NNLO one has
\begin{align}
\tilde{\mathbb C}^{(1)}\big(x,y\big)&=-\mathbb C^{(1)}\big(x,y\big),\\
\tilde{\mathbb C}^{(2)}\big(x,y\big)&=-\mathbb C^{(2)}\big(x,y\big)-\int_{-\infty}^\infty dz\, \tilde{\mathbb C}^{(1)}\big(x,z\big)\mathbb C^{(1)}\big(z,y\big),\nn
\end{align}
where we have suppressed the $\mu/P^z$ dependence in the coefficients for notational simplicity.}

\begin{figure}[htbp]
    \centering
    \subfigure[]{
        \includegraphics[width=0.14\textwidth]{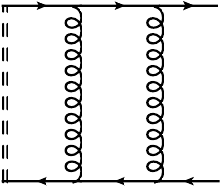}
    }
    \subfigure[]{
        \includegraphics[width=0.139\textwidth]{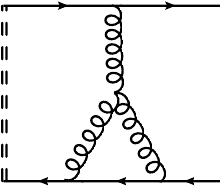}
    }
    \subfigure[]{
        \includegraphics[width=0.14\textwidth]{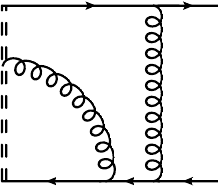}
    }
    \subfigure[]{
        \includegraphics[width=0.14\textwidth]{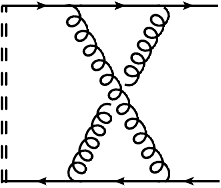}
    }
    \subfigure[]{
        \includegraphics[width=0.14\textwidth]{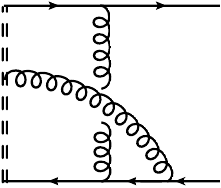}
    }
    \subfigure[]{
        \includegraphics[width=0.14\textwidth]{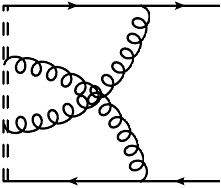}
    }
    \caption{Representative Feynman diagrams for the NNLO calculation of the quasi-LF operator, where the double dashed lines represent the Wilson line. 
    }\label{fig:nnlodiag}
\end{figure}

{\it Perturbative results at NNLO:} 
To extract the NNLO matching kernel, we need to calculate the %quasi-LF correlation in the partonic state at the two-loop order. 
{two-loop QCD corrections to the leading-twist quasi-LF operator $O_q$.}
Our calculation begins in coordinate space, adopting the Feynman gauge. %{\color{blue} adopting the Ioffe-time quasidistribution formalism~\cite{Ioffe:1969kf,Braun:1994jq,Orginos:2017kos}.  
Throughout the calculation, we employ dimensional regularization with $d=4-2\epsilon$ to regularize both UV and infrared (IR) divergences. %where the number of Feynman diagrams is significantly smaller than that in momentum space. 
The momentum space matching kernel is obtained by performing the Fourier transform in Eq.~(\ref{eq:FT_kernel}). %The number of Feynman diagrams in coordinate space is significantly fewer than in momentum space. 
In Fig.~\ref{fig:nnlodiag}, we show some representative Feynman diagrams for the NNLO corrections to the quasi-LF operator.
The first row illustrates typical planar diagrams, %which are more frequently encountered, 
whereas the second row displays non-planar ones. 
 
To illustrate the computational procedure, we take diagram (e) in Fig.~\ref{fig:nnlodiag} as an example. The necessary loop integral takes the following form,
\begin{align}\label{eq:inte}
 I^{(e)}&\!=\!-ig^4 C_F \!\left(\! C_F \!-\! \frac{C_A}{2}\! \right)\! \int_0^1 du \int \frac{d^d l_1 d^d l_2}{(2\pi)^{2d}} 
\, e^{-i(p_2 - l_1 -\bar u l_2) \cdot z}  \notag\\
& \quad \times\frac{ \gamma_\rho (\slashed{l_{12}}-\slashed{p_2})\gamma^{z}\gamma_5 (\slashed{p_1} + \slashed{l_1})\gamma^\rho(\slashed{p_1} - \slashed{l_2} )\slashed{z}}{(l_{12}-p_2)^2  l_2^2(p_1 + l_1)^2(p_1 - l_2)^2 l_2^2l_{12}^2},
\end{align}
where  %$l_{12}=l_1+l_2$, $l_1$ and $l_2$ denote the loop momenta, and $p_1$ and $p_2=(1-y)P$ represent the momenta of external quarks.
{$l_{12}=l_1+l_2$ with $l_1$ and $l_2$ denoting the loop momenta. $p_1$ and $p_2$ represent arbitrary quark momenta introduced to simplify loop integrals in coordinate space. They do not appear in the final result after a trivial Fourier transform that we omit in Eq.~\eqref{eq:inte} for brevity.} To facilitate the calculation, we make the following substitutions: $l_1^\alpha \rightarrow -i \partial_{{{s_1}}}^\alpha\, , l_2^\alpha \rightarrow -i \partial_{{{s_2}}}^\alpha$ together with $ z\to s_1, \, \bar u z \to s_2$. 
In this way, the calculation of tensor integrals in Eq.~(\ref{eq:inte}) reduces to the calculation of scalar integrals supplemented with momentum derivatives w.r.t. $s_1, s_2$. The scalar integral arising from the above equation reads
\begin{align}
J_{12}=\int \frac{d^d l_1 d^d l_2}{(2\pi)^{2d}} \frac{e^{i(l_1 \cdot s_1 + l_2 \cdot s_2)}}{(p_1 - l_2)^2 (p_1 + l_1)^2 (-p_2 + l_{12})^2 l_2^2 l_{12}^2}.
\end{align}
To evaluate such integrals, we apply several formulas collected in the Supplemental Material. After integrating out the loop momenta, the amplitude reduces to an integral over multiple Feynman parameters. By applying appropriate changes of variables and subsequently performing the integration, all Feynman diagram contributions can be turned into a double integral form. Schematically, 
\begin{align}
 \int_0^1 \!\!& d\alpha\, d\beta\, d\gamma\, d\tau\, du \; \{\cdots\} \; e^{i[\cdots]} = \int_0^1 \!\! d\alpha d\beta  e^{{-i (\alpha p_1 + \bar{\beta} p_2 ) \cdot z}} \notag\\
& \times\left\{ 
\theta(1-\alpha-\beta) f_1(\alpha,\beta) + \theta(\alpha+\beta-1) f_2(\alpha,\beta)
\right\},
\end{align}
where the integration variables $\alpha$, $\beta$, $\gamma$, etc., are all Feynman parameters, $ \{\cdots\}, [\cdots]$ denote functions of the Feynman parameters to be integrated, and the final result should retain at most double integrals. 
The integrations over Feynman parameters are partially done using the Mellin-Barnes representation employing \texttt{MB} and \texttt{MBcreate}~\cite{Czakon:2005rk,Belitsky:2022gba}, and some of the resulting hypergeometric functions are $\epsilon$-expanded using~\texttt{HypExp}~\cite{Huber:2005yg}.
Compared to the NLO result, the NNLO result introduces an additional integration region characterized by 
$\theta(\alpha+\beta-1)$, which arises entirely from the non-planar diagrams and has been observed in the evolution kernel for flavor-nonsinglet lightcone operators at NNLO~\cite{Braun:2014vba}. Therefore, $f_1(\alpha,\beta)$ and $f_2(\alpha,\beta)$	
contribute to different kinematic regions of the matching coefficient.

Adding up the contributions from all diagrams, we obtain the bare matching coefficient $\mathbf{C}_B (\alpha, \beta, z^2, \epsilon)$ $\epsilon$-expanded to the necessary order as
\begin{align}
\mathbf{C}_B (\alpha, \beta, z^2, \epsilon) &= \delta(\alpha)\delta(\beta) 
+ a_s^B \left( \frac{1}{\epsilon} \mathbf{C}^{(1)}_{11} 
+ \mathbf{C}^{(1)}_{10} 
+ \epsilon\, \mathbf{C}^{(1)}_{1\mathrm{e}} \right) \notag\\
&\quad + (a_s^B)^2 \left( 
\frac{1}{\epsilon^2} \mathbf{C}^{(2)}_{22} 
+ \frac{1}{\epsilon} \mathbf{C}^{(2)}_{21} 
+ \mathbf{C}^{(2)}_{20} \right)
\end{align}
with $a_s^B$ being the bare coupling constant. 
%where %subscript "B" represents the bare quantities, and 
%$\mathbf{C}_{ij}^{(n)}$ corresponds to the coefficient of $\epsilon$ in the expansion of the bare result. 
The renormalized result in the $\overline{\rm MS}$-scheme is then obtained via
\begin{equation}
C_R(\alpha, \beta, z^2\mu^2) 
= Z_{a_s} Z_{\rm{UV}} 
\left( Z_{O_q}^{l.t.} \otimes \mathbf{C}_{B}(\alpha, \beta, z, \epsilon) \right),
\end{equation}
where $Z_{a_s} = 1-a_s\beta_0/\epsilon + O(a_s^2)$ is the renormalization factor for the strong coupling. %associated with the strong coupling constant.  It absorbs the divergences arising from the scale dependence of the coupling. 
The multiplicative UV renormalization constant $Z_{\rm UV}$ removes the UV singularities and is known to three loops already~\cite{Ji:2015jwa,Braun:2020ymy}. $Z_{O_q}^{l.t.}$ is the renormalization constant for the lightcone operator $O_q^{l.t.}$ in Eq.~\eqref{eq:opefact} extractable from the one-loop and two-loop flavor-nonsinglet evolution kernel given in~\cite{Balitsky:1987bk,Braun:2014vba,Braun:2016qlg}. It removes all IR singularities in the matching coefficient. The fact that the known renormalization constants $Z_{a_s}$, $Z_{\rm UV}$ and $Z_{O_q}^{l.t.}$ exactly remove all $1/{\epsilon^2}$ and $1/\epsilon$ poles from our bare coefficient function $\mathbf{C}_B (\alpha, \beta, z^2, \epsilon) $ provides highly non-trivial checks for our two-loop result. 

Within the $\overline{\rm{MS}}$ scheme, we obtain $C^{\overline{\rm{MS}}}(\alpha, \beta, z^2\mu^2)$ at NNLO level. %The complete analytical expression is given in the Supplemental Material. %At this stage, 
The ratio scheme matching kernel can be obtained by dividing out the corresponding zero-momentum contribution as follows,
\begin{align}\label{eq:ratiomatchingcoord}
C^{\rm{ratio}}(\alpha, \beta, z^2\mu^2)
=\frac{C^{\overline{\rm{MS}}}(\alpha, \beta, z^2\mu^2)}{C^{\overline{\rm{MS}}}_0(z^2\mu^2)},
\end{align}
where {the complete analytical expressions for both $C^{\overline{\rm{MS}}}(\alpha, \beta, z^2\mu^2)$ and $C^{\overline{\rm MS}}_0(z^2\mu^2)$ are given in the Supplemental Material.}
%$C^{\overline{\rm MS}}_0(z^2\mu^2)$ is also given in the Supplementary Material.

As mentioned before, the coordinate space calculation is universal. Therefore, we are able to reproduce the NNLO matching $C^{\overline{\text{MS}}}_{\rm PDF}$ for the quark helicity PDF by taking appropriate kinematic limits, {leading to the following relation} %, and the result is related to $C^{\overline{\rm{MS}}}(\alpha, \beta, z^2\mu^2)$ as
\begin{align}
{C^{\overline{\text{MS}}}_{\rm PDF}}(\alpha', \mu^2 z^2) =
\int_0^1 d\alpha  d\beta\;
\delta(\alpha' - \alpha - \beta)\;C^{\overline{\text{MS}}}(\alpha, \beta, \mu^2 z^2).
\end{align}
It turns out to be the same as the unpolarized quark PDF, and agrees exactly with the coordinate space result in Ref.~\cite{Li:2020xml}.  
This provides further strong evidence supporting the correctness of our result. Moreover, we remark that our coordinate space matching kernel {in $\overline{\rm MS}$ scheme} also applies to both flavor-nonsinglet quark unpolarized and helicity generalized parton distributions (GPDs), defined by $O_q$ in~\eqref{eq:quasilfcorrl} with Dirac structures $\Gamma=\gamma^z$ and $\Gamma=\gamma^z\gamma_5$, respectively.
 
To derive the corresponding matching kernel in momentum space, we perform the Fourier transform following Eq.~\eqref{eq:FT_kernel}. For computational convenience, we divide it into two steps:
\begin{align}
\mathscr{C}(t,y,\mu^2 z^2)&=\int \frac{d\lambda}{2\pi}e^{-it\lambda} \int_0^1 d\alpha d\beta e^{{{i ( \alpha \bar{y} + \bar{\beta} y ) \lambda}}} C(\alpha,\beta,\mu^2z^2),\nn\\
%\end{align}
%where we use the shorthand $\bar{\eta}\equiv 1-\eta$ in the text.  
% \begin{align}
% \mathcal{C}^{(2)}(t, y, \mu^2 z^2) 
% &= \begin{cases}
% [f_1(t, y, \mu^2 z^2)]_+, & 0 < y < t < 1 \\
% [f_1(\bar{t}, \bar{y}, \mu^2 z^2)]_+, & 0 <t < y < 1 
% \end{cases}\notag\\
% &+\begin{cases}
% [f_2(t, y, \mu^2 z^2)]_+, & 0 < t<\bar{y} < 1 \\
% [f_2(\bar{t}, \bar{y}, \mu^2 z^2)]_+, & 0 <  \bar{y}<t < 1 
% \end{cases}.
% \end{align}
% Then, the matching coefficient for quasi-DA are related to the $\mathcal{C}(t,y,\mu^2 z^2)$ as
%\begin{align}
\mathbb{C}(x,y,\mu/P_z)&=\int dt \int\frac{d\lambda}{2\pi}e^{-i(x-t)\lambda}  \mathscr{C}\left(t,y,\frac{\mu^2 \lambda^2}{P_z^2}\right).
\end{align}
In contrast to the NLO case, where the support spans four regions, the NNLO result exhibits a richer structure with four additional physical regions, originating from the %$\theta(1-\alpha-\beta)$ 
{$\theta(\alpha+\beta-1)$} domain in coordinate space. The NNLO matching in momentum space takes the following form:
\begin{align}\label{eq:mtch_kernel_mom}
\mathbb{C}_{\rm ratio}^{(2)}(x, y, \mu/P_z) 
&= \begin{cases}
[h_1(x, y, \mu/P_z)]_+, & x<0 < y  < 1 \\
[h_2(x, y,  \mu/P_z)]_+, & 0 < x< y < 1 \\
[h_2(\bar{x}, \bar{y},  \mu/P_z)]_+, & 0 <y < x < 1 \\
[h_1(\bar{x}, \bar{y},  \mu/P_z)]_+, & 0 <  y < 1<x 
\end{cases}\notag\\
&+\begin{cases}
[h_3(x, y, \mu/P_z)]_+, & x<0 < \bar{y}  < 1 \\
[h_4(x, y,  \mu/P_z)]_+, & 0 < x< \bar{y} < 1 \\
[h_4(\bar{x}, \bar{y},  \mu/P_z)]_+, & 0 <\bar{y} < x < 1 \\
[h_3(\bar{x}, \bar{y},  \mu/P_z)]_+, & 0 <  \bar{y} < 1<x 
\end{cases}.
\end{align}
The singularity arising from $x\rightarrow y$ gets regulated by the plus-prescription
\begin{align}
	\int dx \; [f(x,y)]_+ \;g(x)= \int dx \; f(x,y)\left(g(x)-g(y)\right) .
\end{align}
The complete expression of $\mathbb{C}_{\rm ratio}^{(2)}$ is rather lengthy and can be found in the supplemental Mathematica notebook file. %Note that Eq.~(\ref{eq:mtch_kernel_mom}) is the matching kernel in the ratio scheme, and therefore yields a fully regulated plus-distribution. 
Note that the matching kernel in the ratio scheme constitutes a fully regulated plus-distribution. 
The hybrid scheme matching is related to the ratio scheme matching in coordinate space as follows,
\begin{align}\label{eq:hybridmatchingcoord}
C^{\rm hyb}(\alpha, \beta, z^2\mu^2, z^2/z_s^2)&=C^{\rm ratio}(\alpha, \beta, z^2\mu^2)\\
&\times\Big[1+\Big(\frac{C_0^{\overline{\rm MS}}(z^2\mu^2)}{C_0^{\overline{\rm MS}}(z_s^2\mu^2)}-1\Big)\theta(|z|-z_s)\Big]\, ,\nn
\end{align}
{which introduces an additional contribution to the momentum space kernel, and can be obtained by slightly modifying the Fourier transform of $C^{\rm ratio}(\alpha, \beta, z^2\mu^2)$. The results are given in the Supplemental Material as well.}
%This also introduces an additional contribution to the momentum space result, which can be obtained by slightly modifying the Fourier transform of $C^{\rm ratio}(\alpha, \beta, z^2\mu^2)$. The results are given in the Supplemental Material.

%%%%%%%%%%%%%%%%%%%%%%%%%%%%%%%%%%%%%%%%%%%%%%%%%%%%%%%%%%%%%%%%%%%%%%
%%%%%%%%%%%%%%%%%%%%%%%%%%%%%%%%%%%%%%%%%%%%%%%%%%%%%%%%%%%%%%%%%%%%%%
%\section{Conclusion and outlook}
%\label{SEC:conclusion}
%%%%%%%%%%%%%%%%%%%%%%%%%%%%%%%%%%%%%%%%%%%%%%%%%%%%%%%%%%%%%%%%%%%%%%
%%%%%%%%%%%%%%%%%%%%%%%%%%%%%%%%%%%%%%%%%%%%%%%%%%%%%%%%%%%%%%%%%%%%%%

\begin{figure}[!th]
\begin{center}
\includegraphics[width=0.46\textwidth]{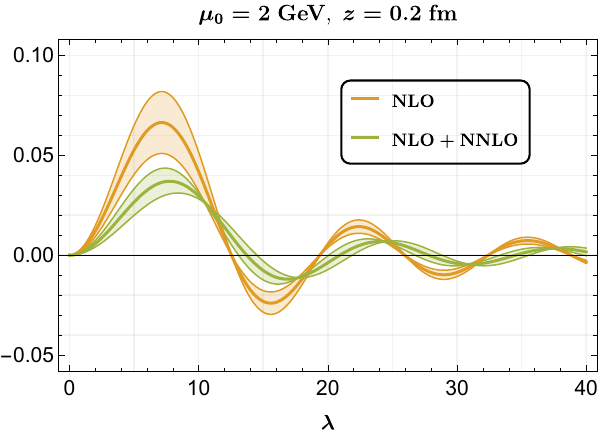} 
\caption{{Higher-order corrections to the coordinate-space DA as a function of $\lambda$ in the ratio scheme, obtained from the asymptotic model.} The orange and green curves indicate the NLO correction and the NLO+NNLO correction, respectively{, with the LO contribution subtracted}. The uncertainty is estimated by varying the renormalization scale $\mu$ between 2 and 4~GeV. 
%\sout{defines the error bars, visualized here as color bands.} 
%{\color{red}is taken symmetrically to define the uncertainty band. The NLO and NLO+NNLO bands are constructed independently, each centered on its own $\mu=2$~GeV curve; the narrower NLO+NNLO band reflects the reduced scale dependence.}
%\sout{The NNLO effect corresponds to roughly $30\%$ of the NLO correction.} 
We have also verified that the evolution effect of the pion LCDA is minuscule up to ${\cal O}(\alpha_s^3)$~\cite{Efremov:1979qk,Lepage:1980fj,Mikhailov:1984ii,Belitsky:1999gu,Braun:2017cih} for $\mu\in(2,4)~{\rm GeV}$.
}\label{fig:nnlonumcoord}
\end{center}
\end{figure}

{\it Numerical impact of NNLO matching:} Now we show the numerical impact of the NNLO matching kernel in both coordinate and momentum space. There have been some lattice determinations of light meson LCDAs based on the short distance factorization approach. However, they all use long-distance correlations where the validity of factorization becomes questionable. 
Therefore, we choose to use the Fourier transform of the asymptotic form $\phi(x)=6x(1-x)$ to illustrate the numerical impact of the NNLO matching kernel in coordinate space 
% Therefore, we have chosen the Fourier transform of the asymptotic form $\phi(x)=6x(1-x)$ as a model to illustrate the numeical impact of the NNLO matching kernel in coordinate space. 
\begin{align}
{h}\left(\alpha,\beta, \lambda\right)
  =\int_0^1 dx\ e^{i x (\alpha - \frac{1}{2})\lambda + i (1 - x) (\frac{1}{2} - \beta)\lambda} \, \phi(x),
\end{align}
where we have introduced the phase factor in such a way that the LF correlation is purely real. With the matching coefficients computed at NLO and NNLO, together with the factorization formula in Eq.~\eqref{eq:coordfact}, we can assess the impact of the NNLO contribution. We set $\mu_{\alpha_s} = \mu_{\rm UV} = \mu_F \equiv \mu$ with $\mu_{\alpha_s}$, $\mu_{\rm UV}$ and $\mu_F$ being the scale for $\alpha_s$, the UV, and the factorization, respectively. We adopt the commonly used $\mu_0=2$~GeV as our default value yielding the central curves in Fig.~\ref{fig:nnlonumcoord}, with $z$ in the matching coefficient fixed at $0.2$~fm. We vary $\mu$ from 2 GeV to 4 GeV, %which reflects the scale dependence of the result  \sout{and is commonly treated as a source of systematic uncertainty.} 
%\sout{As shown in Fig.~\ref{fig:nnlonumcoord}, including NNLO correction significantly reduces the scale dependence, indicating improved perturbative stability. Quantitatively, the NNLO contribution amounts to approximately $30\%$ of the NLO correction.} 
{as shown in Fig.~\ref{fig:nnlonumcoord}, we illustrate the impact of higher-order corrections in coordinate space using the asymptotic distribution amplitude as the input.
The orange and green curves show the NLO and NLO+NNLO corrections, respectively, with the LO contribution subtracted, and the error band comes from the scale variation from $2$ GeV to $4$ GeV. The inclusion of the NNLO correction clearly reduces the
scale dependence, indicating improved perturbative stability.
Note that although the two-loop correction in coordinate space appears sizable, it is moderate in momentum space, as can be seen from Fig.~\ref{fig:nnlonummom} below. The reason is that %it does not imply a large ${\cal O}(\alpha_s^2)$ correction to the momentum-space distribution amplitude.
the latter is obtained through a Fourier transform that integrates over all $\lambda$ and has physical support only on $x\in(0,1)$, so the oscillatory integration tends to average out the coordinate-space corrections.
%This expectation is consistent with the momentum-space extraction in Fig.~\ref{fig:nnlonummom}, where the NNLO correction is found to be moderate. 
}

To demonstrate the numerical impact of the NNLO correction in momentum space, we use the lattice data for the renormalized pion quasi-DA from Ref.~\cite{LatticeParton:2022zqc}, and follow the treatment there to adopt the matching coefficient in Eq.~(\ref{eq:mtch_kernel_mom}). In Fig.~\ref{fig:nnlonummom}, we show the results of the pion LCDA extracted from the lattice data at NLO and NNLO level, respectively. In this analysis, the pion quasi-DA is taken in the continuum limit ($a \to 0$), with $P_z = 2.15$ GeV which is the highest momentum computed in ~\cite{LatticeParton:2022zqc}. %Both the renormalization and factorization scales are set to 2 GeV. 
{Setting $\mu=$ 2 GeV as the default value and varying it from 2 GeV to 4 GeV,} one can see from Fig.~\ref{fig:nnlonummom} that the inclusion of NNLO corrections introduces relatively mild modifications compared to the NLO result, typically amounting to about $10\%$ of the NLO prediction. The uncertainties shown here consist of two parts: the statistical error from the lattice data and the systematic uncertainty associated with varying %the renormalization scale 
$\mu$ from 2 GeV to 4 GeV. The NNLO band is slightly narrower, indicating a reduction in theoretical uncertainties. %Although the corrections are noticeable near the endpoints, 
In the figure, we have shaded the endpoint regions $x<0.1$ and $x>0.9$ where LaMET prediction shall not be trusted. In the Supplemental Material, we also show the matching for two other momenta for comparison purposes, where a similar behavior of the NNLO correction can be observed. 

\begin{figure}[!th]
\begin{center}
\includegraphics[width=0.45\textwidth]{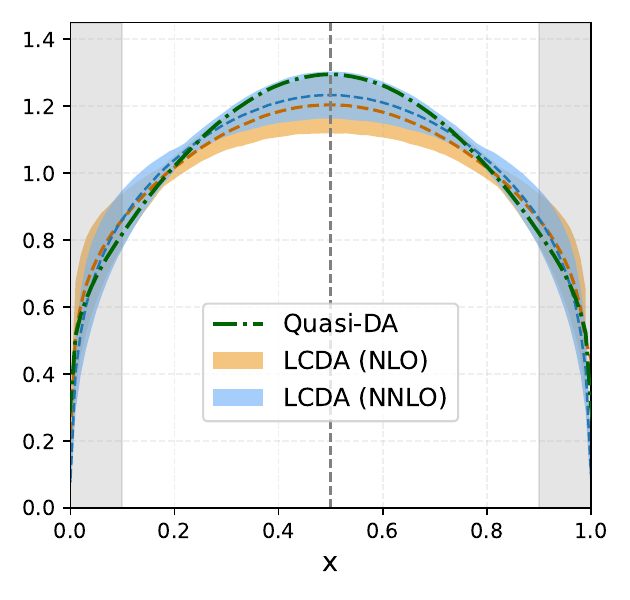} 
\caption{{Extraction of the pion LCDA from lattice data~\cite{LatticeParton:2022zqc} up to the NNLO matching accuracy in momentum space}. The dark green dash-dotted curve represents the continuum-extrapolated quasi-DA at $a \to 0$, $P_z=2.15$ GeV. The orange and blue bands correspond to the LCDA results at NLO and NNLO level, respectively, with the dashed lines indicating their central values.}\label{fig:nnlonummom}
\end{center}
\end{figure}

{\it Summary:} To summarize, we have presented the first complete result of the NNLO matching kernel for the extraction of light meson LCDAs from lattice calculations of quasi-LF correlations. The results are given in both coordinate and momentum space, and thus can be applied to both short-distance factorization and LaMET approaches. {The explicit expressions for the coefficient function in momentum space are available in the \texttt{Mathematica} notebook supplemented to this paper.} {We have applied our result to extract the pion LCDA from lattice data to the NNLO matching accuracy for the first time. The result demonstrates a good convergence of the perturbative matching and establishes the perturbative stability of the LCDA extraction at the two-loop level.} Our coordinate space results can also be directly applied to lattice extraction of flavor-nonsinglet quark unpolarized and helicity GPDs defined by quark bilinear operators with Dirac structures $\Gamma=\gamma^z$ and $\Gamma=\gamma^z\gamma_5$, respectively.

\vspace{2em}

%%%%%%%%%%%%%%%%%%%%%%%%%%%%%%%%%%%%%%%%%%%%%%%%%%%%%%%%%%%%%%%%%%%%%%
%%%%%%%%%%%%%%%%%%%%%%%%%%%%%%%%%%%%%%%%%%%%%%%%%%%%%%%%%%%%%%%%%%%%%%
\acknowledgments
%%%%%%%%%%%%%%%%%%%%%%%%%%%%%%%%%%%%%%%%%%%%%%%%%%%%%%%%%%%%%%%%%%%%%%
%%%%%%%%%%%%%%%%%%%%%%%%%%%%%%%%%%%%%%%%%%%%%%%%%%%%%%%%%%%%%%%%%%%%%%

We thank Jinchen He, Jun Hua, and the LPC collaboration for providing the lattice data used for numerical tests. We also thank Tobias Huber for valuable discussions, and Zhengyang Li and Yanqing Ma for sharing their numerical results of the NNLO isovector unpolarized quark PDF matching with us. YJ acknowledges the support of the DFG grant SFB TRR 257 and the University Development Fund of the Chinese University of Hong Kong, Shenzhen under the grant No.~UDF01003869. FY is partially supported by the U.S. Department of Energy, Office of Science, Office of Nuclear Physics through Contract No. DE-SC0012704, and within the framework of Scientific Discovery through Advanced Computing (SciDAC) award Fundamental Nuclear Physics at the Exascale and Beyond. JHZ is supported in part by National Natural Science Foundation of China under grants No. 12375080, 11975051, 12061131006, the Ministry of Science and Technology of China under Grant No. 2024YFA1611004, and by CUHK-Shenzhen under grant No. UDF01002851.

\bibliographystyle{apsrev}
\bibliography{dannlo}

\newpage

\appendix

\begin{widetext}
\section*{Supplemental Material}\label{sec:supp}

\subsection{Loop integrals in coordinate space}

%The following are some formulas frequently used in the calculation:
We provide some useful formulas for coordinate space calculation with $z^2\neq0$ in this section. In the simplest case, such as the self-energy diagram of a Wilson line, the loop momentum does not appear in the exponent. The corresponding loop integral is thus a standard one. In other cases, such as those in which the exponent contains only one loop momentum (see the ladder diagram (a) in Fig.1 of the main text), the corresponding loop integral %starting from the amplitude 
takes the following form 
\begin{align}
I^{(a)}=\int\frac{d^dl_1d^dl_2}{(2\pi)^{2d}}\,\frac{e^{-iz\cdot(p_2+l_1)}}{{(p_2+l_2)^2 (p_2+l_1)^2 (p_1-l_1)^2 (p_1-l_2)^2 (l_1-l_2)^2 l_2^2}}\, .
\end{align}
It is then convenient to first perform the standard integration over $l_2$, then apply the following formula, 
\begin{align}
    &\int \frac{d^d l}{(2\pi)^d} \, \frac{e^{-i l \cdot z}}{[l^2]^n} 
    = \frac{(-1)^n i}{4^{n - d/2} (4\pi)^{d/2}} \frac{\Gamma(d/2 - n)}{\Gamma(n)} (-z^2)^{n - d/2}\, . 
\end{align}
The same formula also applies to cases where the exponent involves two loop momenta that can be fully decoupled, as in Fig.~1 (f).
However, in some more complicated situations, such as the example presented in the main text (Figure 1(e)), where $l_1$ and $l_2$ cannot be decoupled completely. In this case, without loss of generality, we fix the $d$-dimensional spacelike vector $z^\mu$ along the $z$-direction as $z=(0,\cdots,{\rm z})$. After integrating over the transverse directions (orthogonal to $z$), we then evaluate the integrals over $l_1^z,l_2^z$ using the following formulas,
\begin{align}
&\int \frac{d^d l}{(2\pi)^d} \, \frac{e^{-i l \cdot z}}{[l^2]^n} 
= \frac{(-1)^n i \, \Gamma(n - d/2 + 1/2)}{(4\pi)^{d/2 - 1/2} \Gamma(n)} 
\int \frac{d l^z}{2\pi} \, e^{-i l^z\, {\rm z}} \left( \frac{1}{l_z^2} \right)^{n - d/2 + 1/2}\, ,\notag\\
&\int \frac{d l_1^z d l_2^z}{(2\pi)^2} \, \frac{e^{-i a_1 l_1^z + i a_2 l_2^z}}{\left[ (l_1^z)^2 + (l_2^z)^2 \right]^{n - d}} 
= \frac{\Gamma(-n + d + 1)}{4^{n - d} \pi \Gamma(n - d)} \left[ a_1^2 + a_2^2 \right]^{n - d - 1}\, ,
\end{align}
 giving us, 
\begin{align}
    	I^{(e)}&= \int\frac{d^dl_1d^dl_2}{(2\pi)^{2d}}\,\frac{e^{i(l_1\cdot {s}_1+ l_2\cdot {s}_2)}}{(p_1-l_2)^2 (p_1+l_1)^2 (l_{12}-p_2)^2l_2^2(l_1+l_2)^2}
    	\notag\\
    	&=\frac{\Gamma(-1-2\epsilon)}{4^{1+2\epsilon}(4\pi)^d}\int^1_0 \frac{d\alpha d\beta d\gamma d\tau\,\alpha\tau^{\epsilon}\bar\tau}{(\alpha\bar\alpha)^{1+\epsilon}}e^{-i(\bar\alpha p_1-\alpha\beta p_2)\cdot {s}_1+i[(\tau+\bar\tau\gamma)p_1+\beta\tau p_2]\cdot ({s}_2-\alpha {s}_1)}[-{s}_1^2\alpha\bar\alpha/\tau-({s}_2-\alpha {s}_1)^2]^{1+2\epsilon}\, .
    \end{align}
Note that $s_1$ and $s_2$ should be retained throughout the calculation and only make substitutions $s_1 \to z, \, s_2 \to \bar u z $ after the partial derivatives are taken, as explained in the main text. The above techniques can resolve all issues related to the loop integrals. However, the most challenging part of the computation lies in integrating out the redundant Feynman parameters, which requires numerous changes of integration variables as the starting point.

\subsection{The complete matching coefficient function in coordinate space}

The complete {$\overline{\rm MS}$} scheme coefficient at NNLO level in coordinate space includes two regions, as shown in Eq.~(\ref{eq:opefact}). Its explicit expression reads:
\begin{align}
 C^{\overline{\text{MS}}}(\alpha, \beta, \mu^2 z^2) &= \delta(\alpha)\delta(\beta)
+ a_s(\mu) \left[ {C_{11}(\alpha,\beta) L + C_{10}(\alpha,\beta)} \right]
+ a_s^2(\mu) \left( {C_{22}(\alpha,\beta) L^2 + C_{21}(\alpha,\beta) L + C_{20}(\alpha,\beta)} \right), 
% \widetilde{{\mathcal{C}}}_{\overline{\text{MS}}}(\alpha, \beta, \mu^2 z^2) &= 
% a_s^2(\mu) \left({{\widetilde{C}}_{21} L + {\widetilde{C}}_{20}} \right),
\end{align}
where $L = \ln\left(\frac{-\mu^2 z^2 e^{2 \gamma_E}}{4}\right)$. The NLO expression provided in Ref.~\cite{Yao:2022vtp} reads, %\footnote{$\bar{\alpha}=1-\alpha$, $\bar{\beta}=1-\beta$, $ \widetilde{\alpha\beta}=1-\alpha-\beta$, $\displaystyle\tau=\frac{\alpha\beta}{\bar\alpha\bar\beta}$, $\bar\tau=1-\tau$.}
\begin{align}
&C_{11}(\alpha,\beta)=-2\;C_F\left[1+\left(\frac{\bar{\alpha}}{\alpha}\right)_+ \delta(\beta)+\left(\frac{\bar{\beta}}{\beta}\right)_+\delta(\alpha)-2\delta(\alpha)\delta(\beta)  \right]\theta(1-\alpha-\beta),\\
&C_{10}(\alpha,\beta)= 2\;C_F\left[-\left(\left(\frac{\bar{\alpha}}{\alpha}\right)_+ \delta(\beta)+\left(\frac{\bar{\beta}}{\beta}\right)_+\delta(\alpha)-2\delta(\alpha)\delta(\beta) \right)  +3 -2\left(\frac{\ln\alpha}{\alpha} \right)_+\delta(\beta)-2\left(\frac{\ln\beta}{\beta} \right)_+\delta(\alpha)\right]\theta(1-\alpha-\beta)\, ,
\end{align}
% \begin{align}
%    C^{(2)}(\alpha,\beta) = C_{22}(\alpha,\beta)L^2+C_{21}(\alpha,\beta)L+C_{20}(\alpha,\beta) \notag\\
% \end{align}
The NNLO part takes the form,
\begin{align}
C_{22}(\alpha,\beta)  =& C_F^2 
 \frac{4\bar\tau}{\tau}\Big(1-\delta(\alpha)-\delta(\beta)+\delta(\alpha)\delta(\beta) \Big) \theta(1-\alpha-\beta)+\bigg\{ C_F^2 \bigg[ %-\ln (\widetilde{\alpha\beta}) -2 \ln (\bar{\alpha}) \!+\!4 \ln (\alpha) 
{\ln\!\left(\!\frac{\tau^2}{\bar\tau}\!\right)\!}
\!+\!\left(\frac{4 \bar{\alpha}(\ln \alpha \!-\!1)}{\alpha}\right)_+ \delta(\beta)\!+\!4\delta(\alpha)\delta(\beta) \bigg]  \notag\\
&- C_F \beta_0 \bigg[ \frac12+\left(\frac{\bar{\alpha}}{\alpha}\right)_+ \delta(\beta)- \delta(\alpha)\delta(\beta) \bigg] +(\alpha \leftrightarrow \beta)\bigg\} \theta(1-\alpha-\beta) \, ,
\end{align}
{ where $\bar{\alpha}=1-\alpha$, $\bar{\beta}=1-\beta$, $ \widetilde{\alpha\beta}=1-\alpha-\beta$, $\displaystyle\tau=\frac{\alpha\beta}{\bar\alpha\bar\beta}$, $\bar\tau=1-\tau$, and}
% &  
%  +(\alpha \leftrightarrow \beta)\bigg\} O(z_{12}^{\alpha},z_{21}^{\beta})+ C_F^2%\frac{4 \widetilde{\alpha\beta}}{\alpha\beta} 
%  \frac{4\bar\tau}{\tau}\left( O(z_{12}^{\alpha},z_{21}^{\beta}) - O(z_{12},z_{2}) - O(z_{21},z_{21}^{\beta}) + O(z_{1},z_{2}) \right)\, ,
% \end{align}
\begin{align}
C_{21}(\alpha,\beta) = &\bigg\{ 4C_F\left(C_F\!-\!\frac{C_A}{2}\right) \bigg[{{\Li_2(\tau)-2\Li_2(\bar\alpha)+\frac12\ln\tau\ln\bar\tau+\frac12\ln^2(\bar\alpha\beta)-\ln(\beta\bar\beta)\ln(\widetilde{\alpha\beta})+\left(4-\frac1\tau\right)\ln(\widetilde{\alpha\beta})\!-\!2}} %-\frac{1}{6} \left(12+\pi^2\right)+\ln (\alpha ) (\ln
   %(\alpha )+\ln (\bar{\alpha}))-\ln (\bar{\alpha}) \ln
   %(\widetilde{\alpha\beta})+\ln ^2(\bar{\alpha})-\ln (\alpha )
   %\ln (\widetilde{\alpha\beta})-3
   %\ln (\alpha)
   \notag\\
   &{+\left(\frac2\tau \!-\!\frac1\alpha\!-\!5\right)\ln\bar\alpha -\left(3+\frac{4}{\bar\alpha}\right)\ln\alpha}%-\frac{4\ln (\alpha )}{\bar{\alpha}}-\frac{(6 \beta -\bar{\alpha} (3 \beta +2)) \ln(\bar{\alpha})}{\alpha  \beta }+\frac{(3
   %\alpha  \beta +2 \beta -1) \ln
   %(\widetilde{\alpha\beta})}{\alpha  \beta }+
   %\text{Li}_2\left(-\frac{\widetilde{\alpha\beta}}{\alpha
   %}\right)+\text{Li}_2\left(\frac{\alpha
   %}{\bar{\beta}}\right)
   \bigg]
   +C_F^2\;\bigg[{8\Li_2(\bar\alpha)-8\Li_2(\alpha)+2\ln^2\alpha+\ln^2(\widetilde{\alpha\beta})-4\ln^2\bar\alpha+\frac{8}{\bar\alpha}\ln\alpha} \notag\\%2 \ln ^2(\alpha )\notag\\
   & 
   {+\;6\ln\bar\alpha-8\ln\bar\tau+8} + 4\;\bigg(3\Li_2(\alpha)-\Li_2(\bar\alpha)+\frac{2+\bar\alpha}\alpha\ln^2\alpha+\frac{3\bar\alpha}{2\alpha}\ln\bar\alpha+2\ln\alpha-\frac{2\bar\alpha}\alpha-\frac{\pi^2}{6}\frac{2+\alpha}\alpha\bigg)_+\delta(\beta) \notag\\
   %+\!\bigg(\!-\!\frac{4 \pi ^2 (\alpha +1)}{3 \alpha }\notag\\
&  {-12\; \left(\zeta (3)-\frac{\pi^2}{6}+\frac13\right) \delta(\alpha)\delta(\beta)}\bigg]
  + C_F C_A \bigg[
   {2\Li_2(\alpha)-2\Li_2(\bar\alpha)-\ln^2(\bar\alpha\bar\beta\bar\tau)+\ln^2\bar\alpha+\ln^2\alpha+8\ln\left(\frac{\bar\tau}{\tau}\right)}\notag\\
   &{-10\;\ln\left(\frac{\bar\alpha}\alpha\right)-\frac2\alpha\;\ln\bar\alpha-\frac8{\bar\alpha}\;\ln\alpha-\frac{22}3} %- 2\bigg(8\left( \ln (\bar{\alpha} )
   %\ln (\alpha )+ \text{Li}_2(\bar{\alpha})+
   %\text{Li}_2(\alpha ) \right)-\frac{\left(4 \pi ^2 \alpha \!+\! \left(\pi ^2\!-\!4\right)
   %\bar{\alpha}\right)}{3 \alpha }  \bigg)_+ 
   {\;+\;4\left(\frac{\pi^2}6-\frac23\right)\;\bigg(\frac{\bar\alpha}\alpha\bigg)_+}\;\delta(\beta)
   %\!-\!\left(4 \zeta (3)\!-\!\frac{10}{3}\!+\!2 \pi ^2\!+\!8 \psi
   %^{(2)}(1)\right)
   {\;+\;6\left(\zeta(3)-\frac{\pi^2}6+\frac{5}{18}\right)}\;\delta(\alpha)\delta(\beta) 
   \bigg]\notag\\
   &  %+ C_F \beta_0\; \bigg[ - \ln(\widetilde{\alpha\beta})-\frac{2}{3} -\left(\frac{2 (6 \ln (\alpha )+8 \bar{\alpha}+3 \bar{\alpha}
   %\ln (\bar{\alpha}))}{3 \alpha }\right)_+\delta(\beta)  + \frac{13}{3}\;\delta(\alpha)\delta(\beta)\bigg]
   {- C_F \;\beta_0\; \bigg[  \ln(\widetilde{\alpha\beta})+\frac{2}{3} +2\left(\frac{2}{ \alpha } \ln \alpha +\frac{8\bar{\alpha}}{ 3\alpha }+\frac{\bar{\alpha}
   }{ \alpha }\ln \bar{\alpha}\right)_+\delta(\beta)  - \frac{13}{3}\;\delta(\alpha)\delta(\beta)\bigg]}
   +\;(\alpha \leftrightarrow \beta)
 \bigg\}\;\theta(1-\alpha-\beta)\notag\\
   &  + \bigg\{{4C_F\left(C_F\!-\!\frac{C_A}{2}\right)\bigg[
\Li_2\left(\frac{\bar\alpha}{\beta}\right)-\Li_2\left(\frac{\bar\beta}{\alpha}\right)+\frac12\ln^2\left(-\frac{\bar\tau}{\tau}\right)-\frac{1}{\tau}\ln\left(-\frac{\bar\tau}{\tau}\right)}
\bigg]+ (\alpha \leftrightarrow \beta)\bigg\} \theta(\alpha+\beta-1)\notag\\
   &+C_F^2 \bigg[  %\frac{8 (\bar{\beta} \ln (\alpha ) (\alpha 
   %\beta +\bar{\alpha} \bar{\beta}))}{\alpha  \bar{\alpha} \beta 
   %\bar{\beta}}+\frac{4 \widetilde{\alpha\beta}}{\alpha  \beta }
   {8\left(\frac1{\bar\alpha}+\frac{\bar\beta}{\alpha\beta}\right)\ln\alpha
   +4\frac{\bar\tau}\tau}+ (\alpha \leftrightarrow \beta)\bigg]\Big(1-\delta(\alpha)-\delta(\beta)+\delta(\alpha)\delta(\beta)\Big) \theta(1-\alpha-\beta)\, ,
\end{align}

\begin{align}
   C_{20}(\alpha,\beta) = &\bigg\{2C_F\!\left(\!C_F\!-\!\frac{C_A}{2}\right) \bigg[
     {3\Li_3(\bar\alpha)\!-\!\Li_3(\alpha)+4\Li_3\!\left(\!\frac{\alpha}{\bar\beta}\right)+6\Li_3\!\left(\!\frac{\widetilde{\alpha\beta}}{\bar\beta}\right)\!-\!3\Li_3(\bar\tau)\!-\!\Li_3(\tau)\!+\!\ln\left(\frac{\alpha\bar\alpha\tau^2}{\beta^2\bar\tau^2}\right)\Li_2(\alpha)}\notag\\
    & {-2\;\ln\left(\frac{\bar\alpha}{\widetilde{\alpha\beta}^3}\right)\Li_2\left(\frac\beta{\bar\alpha}\right)-\ln\left(\frac{\beta^2\bar\tau^3}\tau\right)\Li_2(\tau)-2\left(\frac{\alpha-5}{\bar\alpha}+\frac{1+\alpha}{\alpha\beta}\right)\Li_2(\alpha)-2\left(5-\frac2\alpha-\frac{\bar\tau}\tau\right)\Li_2\left(\frac{\beta}{\bar\alpha}\right)}\notag\\
    &- {\frac{13}6\ln^3\alpha+\frac16\ln^3\bar\alpha+\ln\bar\alpha\ln\left(\frac{\tau}{\bar\alpha\bar\beta}\right)\ln\beta+\frac32\;\ln^2\alpha\ln\left(\bar\alpha\bar\beta^2\right)+\frac12\;\ln\left(\bar\alpha\bar\beta\alpha^2\beta^2\right)\ln\tau\ln\bar\tau+\frac52\ln^2\bar\alpha\ln\alpha}\notag\\
    & {-4\ln^2\bar\alpha\ln\bar\beta\!+\!\left(2-\frac1\alpha+\frac{2\alpha}{\bar\beta}\right)\ln(\bar\alpha\bar\tau)\ln\left(\frac{\alpha}{\bar\beta}\right)-\ln\bar\alpha\ln(\alpha\bar\alpha)\ln\left(\frac{\bar\tau}{\tau}\right)-\ln\bar\alpha\ln\bar\beta\ln\tau+\frac12\ln\alpha\ln\beta\ln\left(\frac{\tau}{\bar\tau^2}\right)}\notag\\
    & {+2\ln^2\left(\frac{\alpha}{\bar\beta}\right)\ln(\bar\alpha\beta\bar\tau\tau)-3\ln\left(\frac{\bar\beta}\alpha\right)\ln\left(\frac{\bar\tau}\tau\right)\ln\left(\frac{\bar\alpha\beta}{\bar\beta}\right)\!+\!\left(\frac12\frac\tau{\beta\bar\tau}-\frac{1}{\alpha\beta}-\frac{6}{\alpha\bar\alpha}-\frac\beta\alpha-\frac1\beta-\frac{2\alpha}{\bar\beta}+\frac{17}{2\alpha}\right)\ln^2\alpha}\notag\\
    & {+\frac12\left(\frac{1}{\bar\beta\bar\tau}-\frac{4\beta}{\bar\alpha}-\frac7\alpha+\frac{2+\alpha-2\alpha^2}{\alpha\beta}-9\right)\ln^2\bar\alpha\!-\!\left(1\!+\!\frac{2\bar\alpha}\beta\!-\!\frac{2\beta}{\bar\alpha}\!+\!\frac{1}{\bar\beta\bar\tau}\right)\ln\alpha\ln\bar\alpha\!+\!\left(2\!-\!\frac1\alpha\!+\!\frac{2\alpha}{\bar\beta}\right)\ln\alpha\ln\bar\beta}\notag\\
    & {-\left(\frac{\tau}{2\beta\bar\tau}\!-\!\frac{\bar\beta}{\bar\alpha\tau}\!-\!\frac{1}{2\alpha}\!-\!\frac1\beta\!+\!\frac52\right)\ln\left(\frac{\bar\beta}\alpha\right)\ln\left(\frac{\bar\beta\tau^2}{\alpha\beta^2\bar\tau^2}\right)\!+\!\left(4\!+\!\frac4\alpha\!-\!\frac{2\bar\alpha}\beta\!+\!\frac{2\beta}{\bar\alpha}\!-\!\frac1{\bar\beta\bar\tau}\right)\ln\bar\alpha\ln\left(\beta\frac{\bar\tau}\tau\right)\!+\!\frac{2\pi^2}3\ln\!\left(\!\frac{\bar\tau}\tau\right)}\notag\\
    & {+\left(3\!-\!\frac2\beta\!-\!\frac{2\bar\beta^2}{\alpha\beta}-\frac{2\alpha}{\bar\beta}+\frac{\tau}{\beta\bar\tau}\right)\ln\alpha\ln\left(\beta\frac{\bar\tau}\tau\right)-2\left(\frac{2\pi^2}{3}+\frac1\beta+\frac1\tau-\frac1\alpha\right)\ln\left(\beta\frac{\bar\tau}\tau\right)+2\left(\frac{\alpha-3}\alpha+\frac{2}{\bar\beta\tau}\right)\ln\bar\alpha}\notag\\
    & {-2\;\ln^2\tau\;\ln\bar\tau+\frac16\;\ln^3\tau+\frac2\tau\;\ln\bar\alpha\ln\bar\beta+\left(\frac{\pi^2}6+\frac4\alpha-\frac5{\bar\alpha}-\frac2{\alpha\beta}\right)\ln\alpha+\frac{\pi^2}6\left(11-\frac4{\bar\alpha}-\frac4{\bar\beta}\right)-4(1+\zeta_3) }
    \bigg] \notag\\
   &+C_F^2\bigg[ 
     {8\Li_3(\alpha)\!-\!36\Li_3(\bar\alpha)\!+\!4\left(5\ln\bar\alpha\!+\!4\ln\alpha+\frac{4}{\bar\alpha}\!-\!11\right)\Li_2(\bar\alpha)+\frac16\left(\ln(\widetilde{\alpha\beta})-1\right)^3-\frac13\ln^3\alpha-\frac23\ln^3\bar\alpha}\notag\\
    & {+12\;\ln\bar\alpha\ln^2\alpha+2\;\ln^2\bar\alpha\ln\alpha+\frac{13}2\;\ln^2\bar\alpha-2\left(5-\frac{2}{\bar\alpha}\right)\ln^2\alpha + \frac{16\alpha}{\bar\alpha}\;\ln\bar\alpha\ln\alpha+\left(9\frac{\alpha}{\bar\alpha}-\frac{13}{\bar\alpha}-3\pi^2\right)\ln\alpha}\notag\\    & {+\left(\frac{1}2\!+\!\frac{\pi^2}3\right)\ln(\widetilde{\alpha\beta})\!+\!\left(4\!-\!\frac{\pi^2}3\right)\ln\bar\alpha+32\zeta_3+\frac{11}3\pi^2+\frac{43}6}
    \bigg]
   %&
  \! +\! C_F^2\bigg[ 
    \left(34\!-\!\frac{28}\alpha\right)\Li_3(\bar\alpha) \!+\! \left(\frac{12}\alpha\!-\!14\right)\Li_3(\alpha)\notag\\
    &+\left(26\!-\!\frac{8}\alpha\right)\ln\bar\alpha\Li_2(\alpha)\!+\!\left(10-\frac4\alpha\right)\ln\alpha\Li_2(\alpha)-16\frac{\bar\alpha}\alpha\Li_2(\alpha)+\frac8\alpha\ln^3\alpha-\frac{2\bar\alpha}{3\alpha}\ln^3\bar\alpha-\frac{2-\alpha}\alpha\ln\bar\alpha\ln^2\alpha\notag\\
    &+\;\left(17-\frac6\alpha\right)\;\ln^2\bar\alpha\;\ln\alpha+\left(\frac{7}{2\alpha}+2\alpha-\frac{11}2\right)\;\ln^2\bar\alpha+4\;\left(3-\frac1\alpha-\alpha\right)\;\ln\bar\alpha\;\ln\alpha+2\;\left(\frac6\alpha+\alpha\right)\;\ln^2\alpha\notag\\
    &+\left(\left(10\!+\!\frac{2\pi^2}3\right)\frac{\bar\alpha}\alpha\!-\!3\pi^2\right)\ln\bar\alpha\!+\!\left(7\!-\!\frac{4(3+2\pi^2)}{3\alpha}\right)\ln\alpha\!+\!\left(\frac{20}\alpha\!-\!18\right)\zeta_3\!-\!2\left(\pi^2\left(\alpha\!-\!\frac53\right)\!+\!4\right)\frac{\bar\alpha}\alpha
    \bigg]_+\delta(\beta)\notag\\
    &+{ C_F^2\bigg[\frac{ \pi
   ^4}{45}-42 \zeta (3)+\frac{11 \pi ^2}{3}+10\bigg]\delta(\alpha)\delta(\beta) }
   +C_F C_A\bigg[ 
    7\Li_3(\bar\alpha)-7\Li_3(\alpha)+5\ln\alpha\Li_2(\alpha)-5\ln\bar\alpha\Li_2(\bar\alpha) \notag\\
    &\!-\! 2\left(7\!-\!\frac4{\bar\alpha}\!-\!\frac3\alpha\right)\Li_2(\alpha)\!+\!\frac16\left(\ln^3\alpha\!+\!\ln^3\bar\alpha\!-\!\ln^3(\widetilde{\alpha\beta})\right)\!+\!\frac32\ln\alpha\ln\bar\alpha\ln\left(\frac{\alpha}{\bar\alpha}\right)\! +\! 2\ln^2(\widetilde{\alpha\beta})\!+\!\frac12\left(\frac1\alpha\!-\!7\right)\ln^2\bar\alpha\notag\\
    &-\left(\frac2\alpha+7\right)\ln\bar\alpha\ln\alpha-\frac{13-\alpha}{2\bar\alpha}\ln^2\alpha+\left(\frac{\pi^2}6-\frac{2\bar\alpha}\alpha\right)\ln\bar\alpha+\left(\frac{\pi^2}6-\frac{2+3\alpha}{\bar\alpha}\right)\ln\alpha-\left(\frac{25}6+\frac{\pi^2}6\right)\ln(\widetilde{\alpha\beta})\notag\\
    &-\frac{\pi^2}3\left(\frac{3\alpha}{\bar\alpha}+\frac1{\bar\beta}\right)-\frac{29}{18}
    \bigg]
   \! +\! C_F C_A \bigg[
   2\Li_3(\bar\alpha)\!+\!10\left(1-\frac2\alpha\right)\Li_3(\alpha)\!+\!\frac2\alpha\ln\bar\alpha\Li_2(\alpha)\!-\!2\left(1\!-\!\frac2\alpha\right)\ln\alpha\Li_2(\alpha)\notag\\
   &\!-\!\frac1\alpha\Li_2(\bar\alpha)+16\ln^2\bar\alpha\ln\alpha-15\ln^2\bar\alpha\ln\alpha+\frac\alpha2\ln^2\left(\frac{\alpha}{\bar\alpha}\right)-\frac{8\pi^2}3\ln(\alpha\bar\alpha)-\frac{\ln^2\bar\alpha}{2\alpha}+\left(\frac43+\frac{7\pi^2}3-\frac4{3\alpha}\right)\ln\bar\alpha\notag\\
   &+\left(2+\frac{8\pi^2}3+\frac{4(\pi^2-4)}{3\alpha}\right)\ln\alpha-\left(\frac{76}{18}-\frac{\pi^2}6(1-3\alpha)\right)\frac{\bar\alpha}\alpha-2\left(6-\frac{11}\alpha\right)\zeta_3
   \bigg]_+\delta(\beta) - C_F C_A \bigg[\frac{\pi ^4}{90}\!-\!12\zeta_3 \notag\\
   &+\frac{7 \pi
   ^2}{3}+\frac{22}{9}\bigg]\delta(\alpha)\delta(\beta)  %\displaybreak[1] 
   -C_F\beta_0 \bigg[\frac14 \ln ^2(\widetilde{\alpha\beta})-\frac76
   \ln (\widetilde{\alpha\beta})-\frac{41}{18}
   -\bigg(\frac{4 \Li_2(\alpha
   )}{\alpha }-\frac{4 \ln ^2\alpha}{\alpha }-\frac{\bar{\alpha} \ln
   ^2\bar{\alpha}}{2 \alpha }\!-\!\frac{32 \ln
   \alpha }{3 \alpha }\notag\\
   &\!-\!\frac{11 \bar{\alpha} \ln
   \bar{\alpha}}{3 \alpha }\!-\!\frac{76 \bar{\alpha}+6 \pi
   ^2}{9 \alpha }\bigg)_+\delta(\beta)+ \frac{91}{9}\delta(\alpha)\delta(\beta)\bigg]+\left( \alpha \leftrightarrow \beta \right)\bigg\}\theta(1-\alpha-\beta)+\bigg\{ C_F\left(C_F\!-\!\frac{C_A}{2}\!\right)\!  \bigg[4\Li_3(\alpha)\notag\\
   &    {-12\Li_3\left(-\frac{\widetilde{\alpha\beta}}{\alpha}\right)+8\Li_3\left(-\frac{\widetilde{\alpha\beta}}{\beta}\right)+2\Li_3\left(-\frac{\bar\tau}{\tau}\right)-4\ln\left(\frac{\alpha\bar\tau^2}{\tau^2}\right)\Li_2(\bar\alpha)+4\ln\alpha\Li_2\left(\frac{\bar\alpha}{\beta}\right)-\frac{8\bar\beta}{\beta}\Li_2\left(\frac{\bar\beta}{\alpha}\right)}\notag\\
   &{+4\;(1+\beta)\;\frac{1}{\tau\bar\beta}\;\Li_2(\bar\alpha)+\ln^2\alpha\;\ln\left(\beta\;\bar\beta^2\right)-6\;\ln\bar\alpha\;\ln\alpha\;\ln\left(-\frac{\bar\tau}{\tau}\right)-2\;\ln\left(\frac{\bar\beta}{\alpha}\right)\;\ln\left(-\frac{\beta\bar\tau}{\tau}\right)\ln\left(\frac{\beta\bar\beta^2\bar\tau^4}{\tau^4}\right)}\notag\\
   &{+\ln^2\left(\frac{\bar\alpha}{\beta}\right)\ln\left(-\frac{\bar\alpha\alpha\bar\beta^3\bar\tau^5}{\beta^2}\right)-\ln^2\left(-\frac{\bar\tau}{\tau}\right)\ln\tau+4\ln\left(-\frac{\bar\tau}{\tau}\right)\ln^2\tau+\frac43\ln^3\tau+3\ln\alpha\ln\left(-\widetilde{\alpha\beta}\right)\ln\left(\frac{\bar\alpha}{\beta\bar\beta \tau}\right)}\notag\\
   &{-2\;\ln\tau\;\ln^2\left(-{\bar\beta\bar\tau}\right) - 2\;\ln\left(\frac{\bar\alpha}{\beta}\right)\;\ln\left(-{\bar\beta\bar\tau}\right)\;\ln\left(-\frac{\beta}{\tau\;\bar\tau}\right)+\ln^2\bar\alpha\;\ln\left(\frac{-\beta\;\bar\tau}{\alpha^2\;\tau^5}\right)
   -\frac23\;\ln^3\alpha+\frac{10}3\;\ln^3\left(\;\frac{\alpha}{\bar\alpha}\;\right)}\notag\\
   &{+\left(\frac{8-\alpha+2\alpha\bar\alpha}{\alpha\beta}-\frac{6-\alpha}{\alpha}-\frac{2(2-\alpha)}{\bar\beta}-\frac{3\bar\beta}{\alpha}-\frac{2\alpha}{\widetilde{\alpha\beta}}\right)\ln^2\alpha -2\left(\frac{7\alpha-4}{\alpha\beta}-\frac{5\bar\alpha}{\alpha}\right)\ln\alpha\ln\beta-4\ln^2\left(-\widetilde{\alpha\beta}\right) }\notag\\
   &{+ 2\left(\frac2{\bar\alpha}-\frac2{\alpha}-\frac{1+\bar\alpha^2}{\alpha\beta}\!-\!\frac{\bar\beta}{\alpha}\!+\!\frac{\bar\alpha}{\widetilde{\alpha\beta}}\right) \ln\bar\alpha\ln\beta\!-\!2\left(1+\frac{2-\beta}{\alpha}\!-\!\frac{4-\bar\alpha-\bar\alpha^2}{\alpha\beta}+\frac2{\bar\beta}+\frac{\alpha}{\widetilde{\alpha\beta}}\right)\ln\left(\!-\!\frac{\alpha}{\widetilde{\alpha\beta}}\right)\ln\left(\frac{\bar\beta}{\alpha}\right)}\notag\\
   &{\!+\!\left(4\!+\!\frac{6}{\alpha}\!+\!\frac{2({\bar\alpha}^2\!-\!5)}{\alpha\beta}\!+\!\frac{4\bar\alpha}{\bar\beta}\!+\!\frac{4\bar\beta}{\alpha}\!+\!\frac{2\alpha}{\widetilde{\alpha\beta}}\right)\ln\alpha\ln\left(\!-\!\widetilde{\alpha\beta}\right)\!+\!\left(\!10\!+\!\frac6{\alpha}\!+\!\frac{2\alpha\!+\!4\bar\alpha^2}{\alpha\beta}\!-\!\frac{4\bar\beta}{\bar\alpha}\!+\!\frac{2\bar\beta}{\alpha}\!-\!\frac{2\bar\alpha}{\widetilde{\alpha\beta}}\right)\ln\bar\alpha\!\ln\left(\! \!-\!\widetilde{\alpha\beta}\right)}\notag\\
   &{+ \left(5\!+\!\frac{1}{\alpha\bar\alpha}+\frac{2\beta-1}{\bar\alpha}\!-\!\frac{\bar\alpha}{\beta}\right)\left(\ln\left(\frac{\bar\alpha}{\beta}\right)\ln\left(\frac{\bar\alpha\beta}{\widetilde{\alpha\beta}^2}\right)-\ln^2{\bar\alpha}\right)\!+\!\left(5-\frac{\pi^2}3\!-\!4\frac1\tau\right)\ln\left(\!-\!\widetilde{\alpha\beta}\right)\!+\!\frac{2\pi^2}3\ln\left(\!-\!\frac{\bar\tau}{\tau^2\bar\alpha^2}\right)}\notag\\
   &{\!+\!\frac13\ln^3\left(\!-\!\widetilde{\alpha\beta}\right)+8\ln\alpha\ln\beta\ln\tau\!+\!2\left(\pi^2\!+\!\frac{5\widetilde{\alpha\beta}}{\alpha\beta}\!-\!\frac1\tau\right)\ln\alpha\!+\!\left(1\!+\!\frac{2}{\bar\alpha}\right)\frac1\tau\!-\!\frac1{\alpha\beta}\!-\!2\zeta_3}
   \bigg] \!+\!\left(\alpha \leftrightarrow \beta\right)\bigg\}\theta(\alpha\!+\!\beta\!-\!1)\notag\\
   &+8C_F^2\bigg\{%\frac{8 \ln (\alpha ) \ln (\beta )}{\alpha  \beta}+\frac{8 \ln ^2(\alpha )}{\bar{\alpha}}+\frac{8 \ln(\alpha ) (\alpha  \beta +\bar{\alpha}\bar{\beta})}{\alpha  \bar{\alpha} \beta }+\frac{2\widetilde{\alpha\beta}}{\alpha  \beta } 
   {\frac{\ln \alpha  \ln \beta}{\alpha  \beta} \!+\!\frac{\ln ^2\alpha}{\bar{\alpha}} \!+\!\left( \frac1{\bar\alpha}\!+\!\frac{\bar\beta}{\alpha\beta}\right) \ln\alpha +\frac{\bar\tau}{4\tau}}+\left(\alpha \leftrightarrow \beta\right)\bigg\}
   \Big( 1-\delta(\alpha)\!-\!\delta(\beta)+\delta(\alpha)\delta(\beta)\Big)\theta(1-\alpha-\beta)\, .
\end{align}
We note that the functions entering the two-loop coefficient function have a transcendentality 3, as determined by the presence (absence) of $\Li_3(u),\ln u\Li_2(v),\ln u\ln v\ln w,\cdots$ ($\Li_4(u),\ln u\Li_3(v),\cdots$) with $u,v,w$ being rational functions over either the Feynman parameters $\alpha,\beta$ in coordinate space, or the momentum fractions $x,y$ in momentum space. Whereas the constants have a transcendentality 4 due to the appearance of $\zeta_4 = \pi^4/90$. This is in contrast to the universal transcendentality 4 of the Wilson coefficient function for the Deeply-Virtual Compton Scattering (DVCS) process~\cite{Braun:2020yib,Braun:2021grd,Ji:2023xzk} and the pion-photon transition form factor~\cite{Gao:2021iqq,Braun:2021grd}.

%The quasi-LF correlation function in the ratio scheme $\tilde h^{\rm ratio}_R(z_{12},\lambda=-z_{12}\cdot P)$ is defined by dividing the quasi-LF correlation function $\tilde h(z_{12},\lambda=-z_{12}\cdot P)$ defined in Eq.~\eqref{eq:quasilfcorrl} by its rest frame limit,
% \beq
%\tilde h_R^{\rm ratio}(z_{12},\lambda)=\frac{\tilde h(z_{12},\lambda)}{\tilde h(z_{12},\lambda=0)},
% \eeq
%where the denominator is the quasi-LF correlation at zero momentum. 
To obtain the ratio scheme matching kernel in Eq.~(\ref{eq:ratiomatchingcoord}), we also need the zero momentum contribution $C^{\overline{\rm{MS}}}_0(z^2\mu^2)$, which is given up to ${\cal O}(a_s^2)$ as
%\begin{align}
%C^{\rm{ratio}}(\alpha, \beta, z^2\mu^2)
%=\frac{C^{\overline{\rm{MS}}}(\alpha, \beta, z^2\mu^2)}%{C^{\overline{\rm{MS}}}_0(z^2\mu^2)},
%\end{align}
%
\begin{align}
    C^{\overline{\rm{MS}}}_0(z^2\mu^2) %= C_0(\mu^2z_{12}^2)
    &=\int^1_0d\alpha d\beta\,  C^{\overline{\text{MS}}}(\alpha, \beta, z^2\mu^2 )=1+ a_s (C_{0,11}L+C_{0,10})+a_s^2 (C_{0,22}L^2+C_{0,21}L+C_{0,20}),  \notag\\
    C_{0,11}&=3C_F, \ \ C_{0,10}=7C_F, \ \ C_{0,22}=\frac 9 2 C_F^2+\frac 3 2 C_F\beta_0, \nn\\
    C_{0,21}&=\left(\frac{37}{2}+\frac{8 \pi ^2}{3}\right)C_F^2-\left(1+\frac 2 3 \pi ^2\right)C_F C_A+\frac{19}{2} C_F\beta_0,\nn\\
    C_{0,20}&=\left(-32 \zeta_3+\frac{8 \pi ^2}{3}+\frac{29}{8}\right)C_F^2+\left(8 \zeta_3-3 \pi ^2+\frac{19}{4}\right)C_F C_A+\frac{159}{8} C_F\beta_0, 
%    =&1+a_s C_F\left(3L+7 \right)+a_s^2\bigg\{C_F^2
%   \left(\frac{9 L^2}{2}+\left(\frac{37}{2}+\frac{8 \pi ^2}{3}\right) L-32 \zeta_3+\frac{8 \pi ^2}{3}+\frac{29}{8}\right)\notag\\
%    &+C_F C_A \left(-\frac{1}{3} \left(3+2 \pi ^2\right) L+8 \zeta_3-3 \pi ^2+\frac{19}{4}\right)+\frac{1}{8}
%   C_F \beta_0 \left(12 L^2+76 L+159\right)\bigg\} +{\cal O}(a_s^3)\, ,
\end{align}
where in the first line, we have used the normalization condition of the light meson LCDA $\int^1_0 dx\,\phi(x) = 1$. 

\subsection{The matching coefficient in the hybrid scheme}
The hybrid scheme matching kernel in coordinate space has been given in Eq.~(\ref{eq:hybridmatchingcoord}). Here we derive the result in momentum space.
\begin{align}\label{eq:hybridFT}
{\mathbb C}^{\rm hyb}\Big(x,y,\frac{\mu}{P^z}\Big)&=\int \frac{d\lambda}{2\pi}  e^{-ix\lambda}  \int_0^1 d\alpha d\beta
 \  e^{i({\alpha}+(1-\alpha-\beta)y)\lambda} C^{\rm hyb}\Big(\alpha, \beta, \frac{\lambda^2\mu^2}{P_z^2}, \frac{\lambda^2}{\lambda_s^2}\Big),\nn\\
 &=\int \frac{d\lambda}{2\pi}  e^{-ix\lambda}  \int_0^1 d\alpha d\beta
 \  e^{i({\alpha}+(1-\alpha-\beta)y)\lambda} C^{\rm ratio}\Big(\alpha, \beta, \frac{\lambda^2\mu^2}{P_z^2}\Big)\Big[1+\Big(\frac{C_0^{\overline{\rm MS}}(\lambda^2\mu^2/P_z^2)}{C_0^{\overline{\rm MS}}(\lambda_s^2\mu^2/P_z^2)}-1\Big)\theta(|\lambda|-\lambda_s)\Big],\nn\\
 &=\mathbb{C}^{\rm ratio}\Big(x,y,\frac{\mu}{P_z}\Big)+\delta \mathbb{C}\Big(x,y,\frac{\mu}{P_z}\Big),
\end{align}
where we have made the replacements $z=\lambda/P_z, z_s=\lambda_s/P_z$.

Expand the addtional term $\delta \mathbb{C}\Big(x,y,\frac{\mu}{P_z}\Big)$ to ${\cal O}(a_s^2)$, we have
\begin{align}
\delta \mathbb{C}^{(1)}\Big(x,y,\frac{\mu}{P_z}\Big)&=a_s C_{0,11} {\cal L}_1(y,x,\lambda_s), \nn\\
\delta \mathbb{C}^{(2)}\Big(x,y,\frac{\mu}{P_z}\Big)&=a_s^2[C_{0,22}{\cal L}_2(y,x,\lambda_s)+(2C_{0,22}-C_{0,11}^2)L_s+C_{0,21}-C_{0,10}C_{0,11}{\cal L}_1(y,x,\lambda_s)\nn\\
&+\int_0^1 d\alpha d\beta\, C_{0,11}C_{11}(\alpha,\beta){\cal L}_2(\beta+(1-\alpha-\beta)y,x,\lambda_s)\nn\\
&+\int_0^1 d\alpha d\beta\, C_{0,11}(C_{10}(\alpha,\beta)+C_{11}(\alpha,\beta)L_s){\cal L}_1(\beta+(1-\alpha-\beta)y,x,\lambda_s)],
\end{align}
with $L_s = \ln\left(\frac{\mu^2 z_s^2 e^{2 \gamma_E}}{4}\right)$ and
\begin{align}
{\cal L}_1(u,u_0,\lambda_s)
&=\int\frac{d\lambda}{2\pi}e^{i (u-u_0)\lambda} \ln\frac{\lambda^2}{\lambda_s^2}\theta(|\lambda|-\lambda_s)=\Big[-\frac{1}{|u_0-u|}+\frac{2{\rm Si}((u_0-u)\lambda_s)}{\pi(u_0-u)}\Big]_+\, , \nn\\
{\cal L}_2(u,u_0,\lambda_s)
&=\int\frac{d\lambda}{2\pi}e^{i (u-u_0)\lambda} \ln^2\frac{\lambda^2}{\lambda_s^2}\theta(|\lambda|-\lambda_s)=\Big\{\frac{4}{|u_0-u|}(\ln\lambda_s+\gamma_E+\ln|u_0-u|)\nn\\
&-\frac{4\lambda_s}{\pi}[{}_3 F_3(1,1,1;2,2,2;i(u-u_0)\lambda_s)+{}_3 F_3(1,1,1;2,2,2;-i(u-u_0)\lambda_s)]\Big\}_+\, .
\end{align}

\subsection{The matching effect in momentum space}

In this subsection, we present results using lattice data from Ref.~\cite{LatticeParton:2022zqc} for two additional values of $P_z$ to illustrate the significance of the NNLO contribution in determining the LCDAs. The pion quasi-DAs are at the physical point and extrapolated to the continuum limit ($a \to 0$), with pion momenta $P_z = {1.29, 1.72}$ GeV. We apply the NNLO matching to extract the pion LCDA from the available data, with the central renormalization and factorization scale set to 2 GeV. The uncertainties include both statistical errors and variations due to scale dependence.

\begin{figure*}[h]
  \centering
  \includegraphics[width=0.45\textwidth]{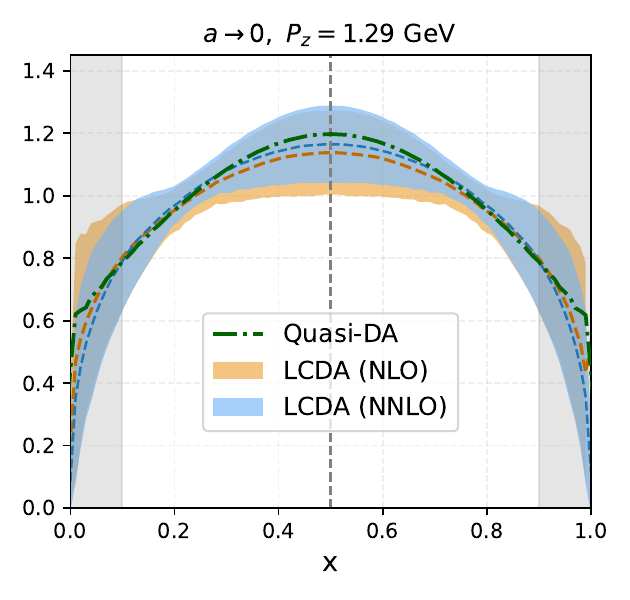}
  \includegraphics[width=0.45\textwidth]{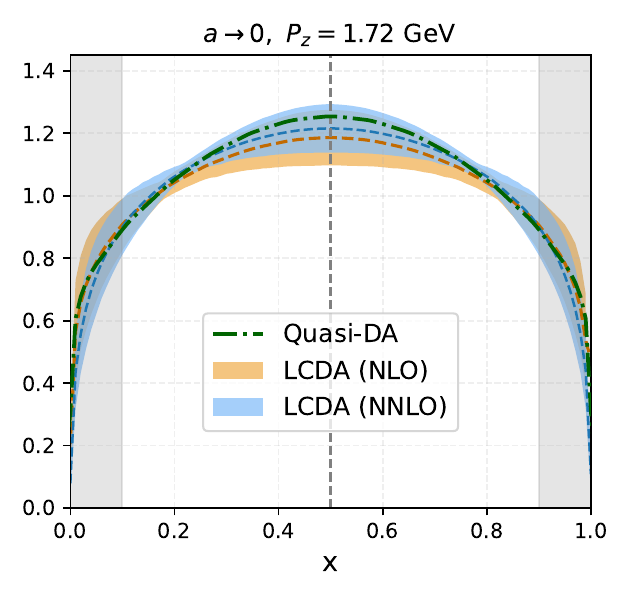}
  \vspace{-1em}
  \caption{Illustration of the numerical impact of the NNLO
matching in momentum space for $P_z =\{ 1.29, 1.72\}$~GeV, similar to Fig.~\ref{fig:nnlonummom} in the main text. Here uncertainties (color bands) come from the statistical error and $\mu$ variation from 2 GeV to 4 GeV.}
  \label{fig:DA_3_mom}
\end{figure*}

As shown in Fig.~\ref{fig:DA_3_mom}, at  $P_z=1.29$~GeV, the NNLO correction is sizeable—contributing roughly $50\%$ of the NLO correction near $x=0.5$. In contrast, for 
$P_z=1.72$~GeV, and for the largest momentum used in the main text ($P_z=2.15$~GeV), the NNLO contribution becomes smaller, indicating that its impact becomes less important with increasing pion momenta.

\end{widetext}

\end{document}